\documentclass[aps,prd,twocolumn,superscriptaddress,nofootinbib,floatfix]{revtex4-2}

\usepackage{amsmath,amssymb,bm}
\usepackage{graphicx}
\usepackage{booktabs}
\usepackage[hidelinks]{hyperref}
\usepackage{xcolor}
\usepackage{microtype}

\graphicspath{{figures/}}

\newcommand{\dd}{\mathrm{d}}
\newcommand{\GeV}{\mathrm{GeV}}
\newcommand{\TeV}{\mathrm{TeV}}
\newcommand{\keV}{\mathrm{keV}}
\newcommand{\cm}{\mathrm{cm}}
\newcommand{\cP}{\mathcal{P}}
\newcommand{\cL}{\mathcal{L}}
\newcommand{\cR}{\mathcal{R}}

\begin{document}

\title{Kaluza--Klein Dark-Photon Mediation of Inelastic Dark Matter at LUX-ZEPLIN}

\author{Waqas Ahmed}
\affiliation{Center for Fundamental Physics, School of Artificial Intelligence, Hubei Polytechnic University, Huangshi 435003, China}
\email{waqasmit@hbpu.edu.cn}
\author{George K. Leontaris}
\email{leonta@uoi.gr}
\affiliation{
Physics Department, Theory Division,
University of Ioannina,
45110 Ioannina, Greece}


\begin{abstract}
{\color{black}
We study a five-dimensional realization of endothermic dark matter motivated by the recent high-recoil LUX-ZEPLIN event. A pseudo-Dirac Standard-Model singlet fermion is localized in the extra dimension and couples to a bulk $U(1)_D$ gauge field. The same localization that suppresses a distant-brane Majorana operator and generates the inelastic splitting also controls the couplings to the Kaluza--Klein (KK) dark-photon tower, producing a localization-filtered mediator propagator. We organize the phenomenology around two representative mass--splitting points: an LZ benchmark, $m_\chi=1.10~\TeV$ and $\delta=365~\keV$, and an IceCube-motivated comparison point, $m_\chi=1.08~\TeV$ and $\delta=566~\keV$. The latter is used only to compare terrestrial and solar kinematics; the published IceCube bound is Higgsino-specific and is not directly imposed on our singlet KK model. At the LZ benchmark with $R^{-1}=0.30~\GeV$ and $m_0=0.10~\GeV$, the KK tower enhances the high-recoil xenon rate by a factor of about four relative to a single dark photon at fixed microscopic coupling. The LZ point lies just below the terrestrial halo speed ceiling, whereas the $566~\keV$ comparison point is inaccessible to ordinary xenon scattering {\color{black}and instead lies in a kinematic regime accessible to solar capture; any neutrino constraint on the present singlet model remains dependent on its capture and annihilation dynamics.} The characteristic prediction of the extra-dimensional construction is therefore a correlated interplay of localization, KK spectral distortion, and inelastic kinematics.
}
\end{abstract}

\maketitle

\section{Introduction}

The microscopic identity of dark matter remains unknown despite increasingly strong constraints from direct-detection, collider, astrophysical, and cosmological searches. Liquid-xenon time-projection chambers have played a leading role in this program, with successive results from XENON1T, PandaX--4T, LZ, and XENONnT pushing conventional WIMP searches to increasingly small cross sections \cite{XENON1T2018,PandaX4T2021,LZExperiment2019,LZFirst2022,XENONnT2023}. At the same time, extending the nuclear-recoil energy window opens sensitivity to less standard interactions whose spectra are weighted toward larger momentum transfer. The LZ Collaboration has recently analyzed nuclear recoils up to approximately $270~\keV$ and reported one event with characteristics consistent with a nuclear recoil near $248~\keV$ in a $2.84$ tonne-year exposure \cite{LZ2026}. The global significance of the departure from the background-only hypothesis is $2.6\sigma$, so the event should not be interpreted as evidence for dark matter. It nevertheless provides a useful theoretical laboratory because an isolated high-energy recoil is qualitatively different from the monotonically falling spectrum expected from standard elastic spin-independent scattering.

Endothermic, or inelastic, dark matter provides one of the simplest mechanisms for shifting spectral weight toward large recoil energy \cite{TuckerSmith2001,TuckerSmith2005,Chang2009,Bramante2016,NagataShirai2015}. If a halo particle $\chi_1$ scatters by transitioning to a slightly heavier state $\chi_2$, the minimum incident speed is controlled by the mass splitting
\begin{equation}
\delta=m_{\chi_2}-m_{\chi_1}>0.
\end{equation}
Splittings of order a few hundred keV can therefore populate the extended high-recoil LZ window. The recently reported event has stimulated a rapidly growing literature exploring endothermic and exothermic inelastic dark matter, Higgsino and other electroweak (EW) realizations, inert- and singlet--doublet models, dark-photon and gauged-baryon-number scenarios, fermionic dark-matter absorption, boosted dark matter, pseudoscalar or axion-like portals, and the associated halo-tail and annual-modulation phenomenology \cite{Su2026,FanReece2026,Freese2026,LouLu2026,DiMauro2026,Yamashita2026,McCabe2026,DuWang2026,Unwin2026,BaerBarger2026,WangXiao2026,Alhazmi2026,DuHuangXie2026,Borah2026,Yuan2026,Langhoff2026}. Given the present single-event statistics, these possibilities should be regarded as alternative theoretical interpretations rather than evidence for a specific microscopic model.

{\color{black}Additional recent analyses of the high-recoil event and inelastic interpretations are given in Refs.~\cite{PalmisanoTammaroTesi2026,NagataYanagida2026}.}

{\color{black}
Several subsequent studies further broaden the range of viable explanations of the high-recoil candidate. Elahi and Schwaller showed that a singlet--doublet Majorana state can generate a hard elastic spin-dependent recoil spectrum along a Higgs blind spot, while Heikinheimo and Zimmermann demonstrated that cosmic-ray-boosted dark matter with momentum-dependent nonrelativistic operators can peak near the observed recoil energy \cite{ElahiSchwaller2026,HeikinheimoZimmermann2026}. These alternatives reinforce the point that the present single event does not by itself select the endothermic interpretation considered here.
}

{\color{black} Solar capture provides an important complementary discriminator. Dark matter falling into the Sun is gravitationally accelerated to velocities substantially larger than those available in terrestrial direct-detection experiments, allowing endothermic transitions with splittings above the xenon kinematic ceiling. A recent analysis of an approximately $1.08 $ TeV Higgsino found that the non-observation of high-energy neutrinos from the Sun by IceCube implies a splitting threshold in the range $\delta\simeq506$--$566~\keV$, depending on the post-capture thermalization treatment \cite{SolarCapture2026,IceCubeSolar2025}. This result excludes the corresponding Higgsino interpretation of the LZ event, but it does not translate directly to the present singlet dark-matter construction. In the Higgsino case solar capture proceeds through tree-level EW scattering on heavy solar elements, and the subsequent annihilation is dominated by $W^+W^{-} $ as well as $ZZ$ final states. In our proposed scenario the dark matter candidate is an SM {\color{black}singlet} field, the capture amplitude is controlled by the localization-filtered dark-photon propagator, while annihilation predominantly proceeds into light dark sector vectors rather than EW gauge bosons. A dedicated calculation of the KK mediated solar capture rate and the resulting neutrino flux would therefore be needed to determine whether the IceCube constraints apply to this scenario. 
 We therefore use the $m_\chi=1.08~\TeV$, $\delta=566~\keV$ point only as an IceCube-motivated comparison benchmark.
}

{\color{black}
Two newer solar-neutrino analyses strengthen the motivation for displaying this comparison while also clarifying its model dependence. Bose \textit{et al.} obtain a Higgsino limit near $\delta\lesssim557~\keV$ using Super-Kamiokande and IceCube data, and Nguyen, Linden, and Hooper confirm that solar-neutrino observations strongly constrain broad classes of inelastic-dark-matter scenarios \cite{BoseEtAl2026,NguyenLindenHooper2026}. We do not transfer these bounds directly to the present singlet KK portal because its solar capture rate and annihilation final states differ from the EW Higgsino case.
}

{\color{black} 
A recurring issue for endothermic interpretations is that a few-hundred-keV splitting can push the required incident speed close to the maximum allowed by the Galactic kinematic boundary. The resulting rate can then depend strongly on the high-velocity tail of the local dark-matter distribution, a sensitivity familiar from earlier studies of astrophysical uncertainties in inelastic direct detection \cite{McCabe2010}. The same kinematics can also produce a pronounced seasonal modulation \cite{McCabe2026,BaerBarger2026,Yuan2026,Langhoff2026}. This motivates treating the overall normalization conservatively and focusing, where possible, on spectral information that is less degenerate with the local halo density and portal strength.

 Dark photons provide a well-studied portal between a hidden sector and the Standard Model through kinetic mixing \cite{Holdom1986,Pospelov2009,Bjorken2009,Essig2013,Alexander2016,Battaglieri2017,DarkPhotonReview}. Of particular relevance, Zhu \textit{et al.} recently studied a four-dimensional endothermic model with a kinetically mixed light dark photon and found TeV-scale dark matter, few-hundred-keV splittings, and a GeV-scale mediator in a combined LZ and relic-density analysis \cite{ZhuEtAl2026}. Our construction is complementary: the mediator is a
bulk field, so the recoil amplitude contains a localization-filtered KK
tower whose residues are correlated with the same geometry that generates
the inelastic splitting.

Extra dimensions bring in additional structure: higher-dimensional fields
appear in four dimensions as Kaluza--Klein towers, with a long history in
large- and universal-extradimensional constructions~\cite{ADD1998,Appelquist2001}.
For a bulk dark photon, the resulting tower of vector mediators can
qualitatively modify annihilation, scattering, and accelerator observables
\cite{Rizzo2018I,Rizzo2018II,RizzoWojcik2021}. Each KK mode contributes a
propagator pole at its own mass, so the momentum dependence of a scattering
or annihilation process is no longer described by a single Breit--Wigner
resonance. In the present context, the relevant momentum transfer is set by
the high-recoil LZ event, which selects a specific window of the KK tower.
}

A complementary geometric realization of the same high-recoil LZ feature was developed in Ref.~\cite{Ahmed:2026qjg}. In that construction, pseudo-Dirac EW doublet dark matter is localized on the visible brane, while dark-number violation is communicated through a neutral bulk state. The exponentially suppressed overlap generates the few-hundred-keV splitting required by endothermic scattering and leads to a characteristic multi-target prediction for xenon and tungsten. The present work addresses a different question. Here the dark matter is a Standard-Model singlet coupled to a bulk $U(1)_D$ gauge field, so the extra dimension enters not only through the origin of the mass splitting but also directly through the mediator propagator. The finite localization width therefore acts as a spectral filter on the KK tower.

In the minimal construction studied below, the Standard Model is localized on one boundary of a flat interval, whereas a $U(1)_D$ gauge field and a vectorlike fermion propagate in the bulk. The fermion zero mode is exponentially localized toward the visible boundary. A dark-number-violating operator on the opposite boundary generates an exponentially small Majorana splitting. Crucially, the finite width of the fermion zero mode also suppresses its overlap with increasingly oscillatory KK gauge modes. Thus the same dimensionless localization parameter $MR$ controls both the inelastic threshold and the KK spectral weights.

The resulting phenomenology is more structured than a simple replacement of one mediator by many. In the strict brane limit the tower sum can be performed analytically, while for finite localization the coupling to the $n$th mode is filtered by an overlap factor $c_n(MR)$. Compactification scales in the sub-GeV to GeV regime are directly relevant because a $250~\keV$ recoil in xenon corresponds to momentum transfer $q\sim0.25~\GeV$. The KK contribution therefore changes not only the normalization but also the recoil-energy dependence. The principal observable of the model is consequently a correlated spectral distortion rather than an isolated enhancement of the total event rate.

The remainder of this paper is organized as follows. In Sec.~II we introduce the five-dimensional dark-sector construction and derive the localization-induced mass splitting and KK couplings. In Sec.~III we construct the localization-filtered KK propagator, while Sec.~IV develops the inelastic xenon-scattering rate and the simplified LZ likelihood. Our numerical results and the resulting parameter correlations are presented in Sec.~V. In Sec.~VI we discuss the main phenomenological implications, theoretical limitations, and possible complementary tests. We summarize our conclusions in Sec.~VII, while technical details of the localization overlap are collected in the Appendix.

\section{Five-dimensional dark sector}

We work on a flat orbifold interval
\begin{equation}
0\le y\le \pi R,
\end{equation}
with compactification scale $M_c=R^{-1}$. The Standard Model fields are localized at $y=0$. The dark sector contains a bulk Abelian gauge field $A_M^D$, two bulk fermions whose orbifold parities leave a vectorlike neutral zero mode, and a dark Higgs field $\Phi$ responsible for breaking $U(1)_D$.

A convenient representative assignment is to take two five-dimensional fermions,  $\Psi_1$ and $\Psi_2$, with the same dark charge
$q_D$ but opposite orbifold parities, so that a left-handed zero mode survives from one field and a right-handed zero mode from the other. Opposite signs of their odd bulk masses may then be chosen so that both surviving zero modes are localized toward $y=0$ with the same exponentially falling profile $f_{\chi}(y) = {\cal N}_{\chi} e^{-M y}$ with ${\cal N}_{\chi}$ fixed by canonical normalization.

A dark-Higgs charge $Q_D(\Phi)=-2q_D$ makes the distant-brane Majorana operator gauge invariant. This is summarized in Table~\ref{tab:fieldcontent}; the low-energy results depend only on the resulting vectorlike zero mode, its localization profile, and the gauge-invariant Majorana insertion.

\begin{table}[t]
\caption{Representative field assignment realizing the effective five-dimensional construction. The entries in the parity column refer to the four-dimensional left- and right-handed components.}
\label{tab:fieldcontent}
\begin{ruledtabular}
\begin{tabular}{lccc}
Field & $U(1)_D$ charge & $(P_L,P_R)$ & zero mode\\
\hline
$\Psi_1$ & $q_D$ & $(+,-)$ & $\chi_L$\\
$\Psi_2$ & $q_D$ & $(-,+)$ & $\chi_R$\\
$\Phi$    & $-2q_D$ & $(+,+)$ & dark Higgs
\end{tabular}
\end{ruledtabular}
\end{table}

A schematic five-dimensional action is
\begin{align}
S=&\int d^4x\int_0^{\pi R}dy\Big[-\frac14F^D_{MN}F_D^{MN}
+|D_M\Phi|^2-V(\Phi)\Big]\nonumber\\
&+\int d^4x\int_0^{\pi R}dy\,
\bar\Psi(i\Gamma^M D_M-M\,\epsilon(y))\Psi
\nonumber\\
&-\int d^4x\,\frac{\epsilon_5}{2\sqrt{\Lambda_5}}
B_{\mu\nu}F_D^{\mu\nu}(x,0)
+S_{\rm Maj},
\label{eq:5daction}
\end{align}
Here $B_{\mu\nu}(x)$  is the four-dimensional hypercharge field strength, which has no dependence on the extra coordinate because the Standard Model is localized at $y=0$. The bulk dark field strength $F_D^{\mu\nu}(x,0)$ is evaluated at the same boundary.
Furthermore,  $\Lambda_5$ is the cutoff and $\epsilon_5$ is a kinetic mixing parameter. Brane-localized kinetic mixing of a bulk dark  photon has been studied in detail in Refs.~\cite{Rizzo2018I,Rizzo2018II,RizzoWojcik2021}. We treat the mixing perturbatively and work with the resulting four-dimensional zero-mode parameter $\epsilon$.
With the mode normalization used below, the brane operator implies at leading
order
\begin{equation}
\epsilon\equiv\epsilon_0
=\frac{\epsilon_5}{\sqrt{\pi R\Lambda_5}},
\qquad
\epsilon_n=\sqrt{2}\,\epsilon
\quad (n\geq1),
\label{eq:mixingmodes}
\end{equation}
where $\epsilon_n$ denotes the effective mixing of the $n^{th} $ KK mode with the
visible current. We regard $\epsilon$ as the renormalized low-energy input
parameter; possible boundary kinetic terms or ultraviolet threshold corrections
can modify the simple relation to $\epsilon_5$ without changing the
localization argument developed below. For compactness, $\Psi$ in
Eq.~\eqref{eq:5daction} denotes the bulk fermion pair schematically.

The canonically normalized gauge zero mode couples to the normalized fermion
zero mode with
\begin{equation}
g_D\equiv g_{\chi0}=\frac{g_5}{\sqrt{\pi R}},
\label{eq:g4g5}
\end{equation}
where $g_5$ is the five-dimensional gauge coupling. This relation fixes the
normalization used below for $\alpha_D=g_D^2/(4\pi)$ and for the relative
excited-mode couplings.

The odd bulk mass localizes the fermion zero mode toward $y=0$,
\begin{equation}
f_\chi(y)=\mathcal N_\chi e^{-My},
\qquad
\mathcal N_\chi^2=\frac{2M}{1-e^{-2\pi MR}}.
\label{eq:fermionprofile}
\end{equation}
Exponential localization is a standard mechanism for generating small effective couplings from higher-dimensional geography \cite{ArkaniHamedSchmaltz2000,ArkaniHamedGrossman2000,GrossmanNeubert2000}.

We assume dark-number violation is localized on the distant boundary. A representative gauge-invariant operator is
\begin{equation}
S_{\rm Maj}\supset
-\int d^4x\int dy\,
\delta(y-\pi R)
\frac{\lambda_M}{\Lambda_5^{3/2}}
\Phi\,\Psi\Psi+\text{H.c.},
\label{eq:majop}
\end{equation}
where the dark Higgs charge is chosen to make the operator $U(1)_D$ invariant before symmetry breaking. After $\Phi$ develops a vacuum expectation value and the zero mode is inserted, the induced Majorana mass inherits two powers of the distant-boundary wave function.

{\color{black}
Inserting the normalized zero-mode profile into the distant-brane operator shows explicitly that the induced Majorana insertion scales as
\begin{equation}
m_M^{\rm eff}\propto
\frac{\lambda_M\langle\Phi\rangle}{\Lambda_5^{3/2}}
\left|f_\chi(\pi R)\right|^2
=\frac{\lambda_M\langle\Phi\rangle}{\Lambda_5^{3/2}}
\mathcal N_\chi^2 e^{-2\pi MR}.
\label{eq:majscaling}
\end{equation}
Thus the exponential suppression follows directly from the zero-mode profile; the remaining normalization and ultraviolet factors can be absorbed into a single low-energy coefficient $\delta_{\rm UV}$.
}
It is convenient to write
\begin{equation}
\delta\equiv m_{\chi_2}-m_{\chi_1}
=\delta_{\rm UV}e^{-2\pi MR},
\label{eq:deltaMR}
\end{equation}
where $\delta_{\rm UV}$ collects the brane coefficient, cutoff dependence, dark-Higgs vacuum expectation value and normalization factors. Equation~\eqref{eq:deltaMR} is an effective low-energy parameterization of the geometric suppression; a UV completion may correlate $\delta_{\rm UV}$ with the gauge-boson mass and other boundary operators.

The inverse relation,
\begin{equation}
MR=\frac{1}{2\pi}\ln\left(\frac{\delta_{\rm UV}}{\delta}\right),
\label{eq:MRinverse}
\end{equation}
makes the inelastic splitting a direct probe of the localization parameter.
{\color{black}For the benchmark adopted here,}
$\delta=365~\keV$ and $\delta_{\rm UV}=1~\GeV$, we obtain
\begin{equation}
MR=1.260.
\end{equation}
The dependence on the ultraviolet coefficient is only logarithmic. For the same
$\delta=365~\keV$, varying $\delta_{\rm UV}$ by one order of magnitude around
the benchmark gives
\begin{equation}
{\color{black}
\begin{aligned}
\delta_{\rm UV}&=(0.1,\ 1,\ 10)~\GeV,\\
&\Longrightarrow MR=(0.893,\ 1.260,\ 1.626),
\end{aligned}
}
\label{eq:MRrobust}
\end{equation}
so the finite-width localization regime is not tied to a sharply tuned value of
$\delta_{\rm UV}$. We therefore do not repeat the geometric-splitting plot here and instead focus
on the additional mediator-sector information generated by the bulk dark
photon. {\color{black}The compact mediator dimension considered here is phenomenologically distinct: its sub-GeV KK spacing is comparable to the momentum transfer in the LZ high-recoil window.}

For Neumann boundary conditions on $A_\mu^D$, the mode expansion is
\begin{equation}
A_\mu^D(x,y)=\frac{1}{\sqrt{\pi R}}A_\mu^{(0)}(x)
+\sqrt{\frac{2}{\pi R}}\sum_{n=1}^\infty A_\mu^{(n)}(x)\cos\frac{ny}{R}.
\label{eq:gaugemode}
\end{equation}
A uniform bulk contribution to the dark-photon mass gives
\begin{equation}
m_n^2=m_0^2+\frac{n^2}{R^2}.
\label{eq:KKmass}
\end{equation}
For the benchmark $m_0=0.10~\GeV$ and $R^{-1}=0.30~\GeV$, Eq.~\eqref{eq:KKmass} gives $m_1\simeq0.316~\GeV$, placing the first KK excitation directly in the momentum-transfer range relevant for the high-energy LZ event.

The coupling of the fermion zero mode to $A_\mu^{(n)}$ is an overlap integral. Relative to the zero-mode coupling,
\begin{equation}
c_n(MR)\equiv\frac{g_{\chi n}}{g_{\chi0}}
=\sqrt2\,
\frac{(2MR)^2\left[1-(-1)^n e^{-2\pi MR}\right]}
{\left[(2MR)^2+n^2\right]\left[1-e^{-2\pi MR}\right]}.
\label{eq:cn}
\end{equation}
In the strict brane limit $MR\rightarrow\infty$, the coupling factors
approach $c_n\rightarrow\sqrt{2}$ for each fixed low-lying KK mode.
For finite localization, however, the increasingly oscillatory KK
wavefunctions are averaged over the smooth fermion profile, leading to a progressive suppression of the higher modes. For the benchmark
$MR=1.260$, the lowest KK modes therefore retain appreciable couplings,
whereas the higher excitations are increasingly filtered. This
localization effect is subsequently encoded in the KK propagator and
produces a momentum-dependent modification of the recoil spectrum.

For the benchmark $MR=1.260$, Eq.~\eqref{eq:cn} gives $c_1\simeq1.22$, $c_2\simeq0.87$, and $c_3\simeq0.59$. Thus the first excitation remains close to the brane-localized value while the higher modes are already visibly filtered, without requiring a separate coupling-spectrum figure. This correlation is the main model-building distinction of the present setup. The parameter $MR$ is not merely responsible for a small mass splitting; it also controls which KK mediators participate in direct detection.

\section{Localization-filtered KK propagator}
{\color{black}
In the previous section we derived the couplings $c_n(M R)$  of the dark-matter zero mode to the individual KK gauge modes. 
The next step is to assemble these couplings into the propagator that mediates dark-matter scattering on a nucleus.
This propagator  encodes both the tower of mediator poles and the localization-induced suppression of their residues. 
To isolate the effect of localization, it is useful to first compute  the propagator in the strict brane limit, 
in which the dark-matter current is also localized at $y=0$.}

Let then define
\begin{equation}
Q\equiv\sqrt{q^2+m_0^2}.
\end{equation}
If the two interaction vertices were exactly localized on the same brane, each excited gauge mode would contribute a factor of two relative to the zero mode and the propagator would be
\begin{equation}
\cP_{\rm brane}(q)=\frac{1}{Q^2}
+2\sum_{n=1}^\infty\frac{1}{Q^2+n^2/R^2}.
\label{eq:Pbranesum}
\end{equation}
Using the standard identity for $\sum_{n=-\infty}^{\infty}(n^2+a^2)^{-1}$, this becomes
\begin{equation}
\cP_{\rm brane}(q)=\frac{\pi R}{Q}\coth(\pi RQ).
\label{eq:Pbrane}
\end{equation}
It is useful to expose the two limiting regimes analytically. Defining
$x\equiv\pi RQ$, the unfiltered brane result satisfies
\begin{equation}
\frac{\cP_{\rm brane}}{\cP_0}=x\coth x.
\label{eq:braneratio}
\end{equation}
Hence
\begin{equation}
x\coth x=
\begin{cases}
1+\dfrac{x^2}{3}+\mathcal O(x^4), & x\ll1,\\[2mm]
x\left[1+2e^{-2x}+\cdots\right], & x\gg1.
\end{cases}
\label{eq:branelimits}
\end{equation}
The first limit corresponds to decoupled excited states, whereas the second
shows the coherent enhancement of a dense unfiltered tower. Finite fermion
localization always reduces this brane-limit envelope by suppressing the
higher-mode residues.

The ordinary four-dimensional propagator is simply
\begin{equation}
\cP_0(q)=\frac{1}{Q^2}.
\end{equation}

In our model the visible current couples to the zero mode with strength $\epsilon e$ and to each excited mode with the brane wave-function factor $\sqrt2\,\epsilon e$, as in Eq.~\eqref{eq:mixingmodes}. The dark-matter vertex is instead weighted by $c_n(MR)$ from Eq.~\eqref{eq:cn}. Normalizing to the zero-mode product of couplings therefore gives
\begin{equation}
\cP_{\rm loc}(q)=\frac{1}{Q^2}
+\sum_{n=1}^\infty
\frac{\sqrt2\,c_n(MR)}{Q^2+n^2/R^2}.
\label{eq:Ploc}
\end{equation}
For numerical evaluation we truncate the mathematical KK sum at $n_{\max}=100$, which is sufficient for convergence of the displayed low-momentum observables. This numerical truncation should not be identified with the physical cutoff of the five-dimensional effective theory; in a UV-complete treatment only modes satisfying $m_n\lesssim\Lambda_5$ should be retained. As a code-level validation, the unfiltered numerical sum reproduces Eq.~\eqref{eq:Pbrane} to better than $4\times10^{-4}$ over $q=0.05$--$0.5~\GeV$. The rapid convergence of the localized tower can also
be understood analytically. For $n\gg2MR$,
\begin{equation}
c_n(MR)\simeq
\sqrt2\,\frac{(2MR)^2}{n^2},
\label{eq:cnasymptotic}
\end{equation}
up to the exponentially small parity-dependent correction. Since the KK
propagator denominator contributes another factor $n^{-2}$ at large $n$, the
individual localized terms fall as $n^{-4}$. The low-momentum observables are
therefore dominated by the first few KK states rather than by the ultraviolet
end of the numerical sum.

\begin{figure*}[t]
\centering
\includegraphics[width=0.94\textwidth]{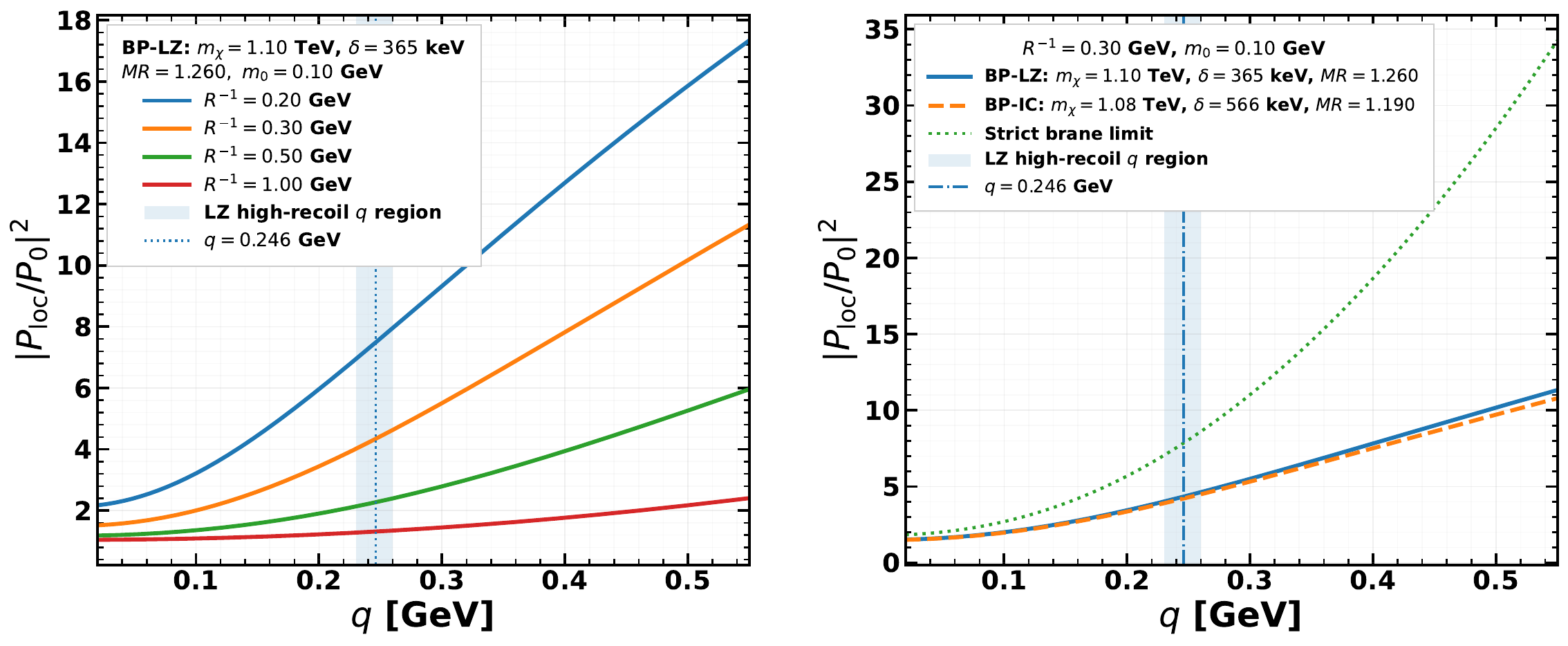}
\caption{Updated propagator comparison. Left: squared localization-filtered KK enhancement $|P_{\rm loc}/P_0|^2$ for BP-LZ as the compactification scale is varied over $R^{-1}=0.20$, $0.30$, $0.50$, and $1.00~\GeV$. Right: comparison of BP-LZ and BP-IC at the common mediator point $R^{-1}=0.30~\GeV$, $m_0=0.10~\GeV$, together with the strict-brane limit. The shaded strip indicates the characteristic momentum-transfer region of the LZ high-recoil event.}
\label{fig:enhancement}
\end{figure*}

The quantity relevant for the scattering rate is $|\cP_{\rm loc}/\cP_0|^2$. Figure~\ref{fig:enhancement} shows that the KK tower becomes important once $qR$ is of order unity or larger. A xenon recoil near $248~\keV$ corresponds to $q\simeq0.24$--$0.25~\GeV$, precisely where compactification scales in the sub-GeV range can produce order-one to multi-fold effects. At the reference point $q=0.246~\GeV$,
$m_0=0.10~\GeV$ and $R^{-1}=0.30~\GeV$, one has $x\simeq2.78$.
The unfiltered brane tower would give
$|\cP_{\rm brane}/\cP_0|^2\simeq7.85$, whereas the finite-width profile with
$MR=1.260$ reduces this to
$|\cP_{\rm loc}/\cP_0|^2\simeq4.34$. This comparison makes the role of
localization especially transparent: the first KK modes remain relevant, but
the tower is significantly filtered before reaching the strict brane limit.

Table~\ref{tab:kkmodes} shows explicitly why only the first few states matter
at the benchmark. The final column gives the contribution of an individual
excited mode to the propagator amplitude, normalized to the zero-mode
amplitude, at $q=0.246~\GeV$.
\begin{table}[t]
\caption{Localization filtering of the first KK modes for
$MR=1.260$, $R^{-1}=0.30~\GeV$ and $m_0=0.10~\GeV$.}
\label{tab:kkmodes}
\begin{ruledtabular}
\begin{tabular}{cccc}
$n$ & $m_n$ [GeV] & $c_n$ &
$\Delta\cP_n/\cP_0$\\
\hline
1 & 0.316 & 1.223 & 0.760\\
2 & 0.608 & 0.868 & 0.201\\
3 & 0.906 & 0.585 & 0.066\\
4 & 1.204 & 0.402 & 0.027\\
5 & 1.503 & 0.287 & 0.012
\end{tabular}
\end{ruledtabular}
\end{table}

\section{Inelastic xenon scattering}

The scattering process is
\begin{equation}
\chi_1+N\rightarrow\chi_2+N,
\end{equation}
with positive splitting $\delta$. The minimum incoming speed needed to produce recoil energy $E_R$ is \cite{TuckerSmith2001,TuckerSmith2005}
\begin{equation}
v_{\min}(E_R)=\frac{m_AE_R/\mu_{\chi A}+\delta}
{\sqrt{2m_AE_R}},
\label{eq:vmin}
\end{equation}
where $\mu_{\chi A}=m_\chi m_A/(m_\chi+m_A)$ is the reduced mass. Large splittings select the fastest part of the halo and can confine events to a relatively narrow high-energy interval. The minimum of Eq.~\eqref{eq:vmin} occurs at
\begin{equation}
E_R^\star=\frac{\delta\,\mu_{\chi A}}{m_A},
\qquad
v_{\min}^\star=\sqrt{\frac{2\delta}{\mu_{\chi A}}},
\label{eq:kinematicminimum}
\end{equation}
which makes the kinematic origin of the high-recoil preference transparent.
{\color{black}
For a representative $^{131}\mathrm{Xe}$ nucleus, the LZ benchmark
$m_\chi=1.10~\TeV$ and $\delta=365~\keV$ gives
$E_R^\star\simeq329~\keV$ and $v_{\min}^\star\simeq773~\mathrm{km/s}$.
At $E_R=248~\keV$ one finds
$v_{\min}\simeq781~\mathrm{km/s}$, only slightly below
$v_{\rm esc}+v_E\simeq795~\mathrm{km/s}$. By contrast, at the
IceCube-motivated comparison point
$m_\chi=1.08~\TeV$ and $\delta=566~\keV$ the same recoil requires
$v_{\min}(248~\keV)\simeq1.03\times10^3~\mathrm{km/s}$, well above the
terrestrial halo ceiling. The comparison therefore separates two physically
distinct regimes: the LZ point probes the extreme Galactic tail, whereas the
larger splitting requires the additional gravitational acceleration available
inside the Sun.
} Our numerical implementation uses a truncated Maxwellian Standard Halo Model with
\begin{align}
\rho_\chi&=0.3~\GeV\,\cm^{-3},\qquad
v_0=238~\mathrm{km/s},\nonumber\\
v_{\rm esc}&=544~\mathrm{km/s}.
\end{align}
and a representative Earth speed $v_E=250.5~\mathrm{km/s}$, following common direct-detection conventions \cite{LewinSmith1996,Baxter2021}. The corresponding kinematic thresholds are evaluated directly from Eq.~\eqref{eq:vmin} in the numerical rate calculation. In the Galactic frame we use
{\color{black}
\begin{align}
f_G(\bm v)&=\frac{e^{-v^2/v_0^2}}
{N_{\rm esc}\,\pi^{3/2}v_0^3}\,
\Theta(v_{\rm esc}-v),\nonumber\\
N_{\rm esc}&=\operatorname{erf}(z)
-\frac{2z}{\sqrt{\pi}}e^{-z^2},
\label{eq:shm}
\end{align}
}
with $z=v_{\rm esc}/v_0$, and boost to the Earth frame through
$f_E(\bm v)=f_G(\bm v+\bm v_E)$. It is useful to define the mean inverse
speed
\begin{equation}
\eta(v_{\min})\equiv
\int_{v>v_{\min}}\frac{\dd^3v}{v}\,f_E(\bm v),
\label{eq:eta}
\end{equation}
which contains all of the halo dependence of the leading vector-mediated
rate.

After diagonalizing the small Majorana terms, the vector current of a pseudo-Dirac fermion is dominantly off diagonal. Kinetic mixing therefore mediates endothermic scattering without requiring a diagonal elastic vector interaction at leading order. Defining
\begin{equation}
\alpha_D\equiv\frac{g_D^2}{4\pi},
\end{equation}
the differential cross section for a nucleus of charge $Z$ is
\begin{equation}
\frac{\dd\sigma_A}{\dd E_R}
=\frac{m_A}{2\pi v^2}
\left[g_D\epsilon e Z\cP(q)\right]^2
F_A^2(q),
\label{eq:dsigma}
\end{equation}
with $q^2=2m_AE_R$. For the four-dimensional comparison $\cP=\cP_0$; for the five-dimensional model $\cP=\cP_{\rm loc}$. Equation~\eqref{eq:dsigma} is equivalent to the usual vector-portal expression $8\pi\alpha\alpha_D\epsilon^2 Z^2m_A/[v^2(q^2+m_{A'}^2)^2]$ when only a single mediator is retained.

For a transparent baseline we use the standard Helm form factor \cite{Helm1956,LewinSmith1996}; more general nonrelativistic nuclear responses can be organized systematically in the effective-operator framework \cite{Fitzpatrick2013,Anand2014}. Thus,
\begin{equation}
F_A(q)=3\,\frac{j_1(qr_n)}{qr_n}
\exp\!\left[-\frac{(qs)^2}{2}\right],
\label{eq:helm}
\end{equation}
with
\begin{equation}
r_n^2=c^2+\frac{7\pi^2a^2}{3}-5s^2,\qquad
c=1.23A^{1/3}-0.60~\mathrm{fm},
\label{eq:helmradius}
\end{equation}
and $a=0.52~\mathrm{fm}$, $s=0.90~\mathrm{fm}$. We sum over natural xenon
isotopes with their atom fractions. \footnote{The role of xenon-isotope filtering near the inelastic
kinematic edge has also been discussed in
Ref.~\cite{AhmedAhmadRehman2026}}This provides a transparent and
reproducible baseline comparable to simple high-energy inelastic-DM analyses.
A full detector-level LZ recast should ultimately use the collaboration's
detailed nuclear response and detector simulation rather than the Helm
approximation alone.

The event rate is
\begin{equation}
\frac{\dd R}{\dd E_R}=N_T\frac{\rho_\chi}{m_\chi}
\int_{v>v_{\min}}\dd^3v\,f_E(\bm v)\,v
\frac{\dd\sigma_A}{\dd E_R}.
\label{eq:rate}
\end{equation}
Using Eq.~\eqref{eq:eta}, the rate for a single isotope can equivalently be
written as
\begin{equation}
\frac{\dd R_A}{\dd E_R}
=N_T\frac{\rho_\chi}{m_\chi}\,
\frac{m_A}{2\pi}
\left[g_D\epsilon e Z\cP(q)\right]^2
F_A^2(q)\,\eta(v_{\min}),
\label{eq:rateeta}
\end{equation}
which makes the separation between particle physics, nuclear response and halo
kinematics explicit. All rates quoted below include the natural-xenon isotope
sum with atom-fraction normalization.

To characterize the high-energy feature without claiming to reproduce the full LZ detector-level profile likelihood, we follow the illustrative statistical construction of Ref.~\cite{Su2026}. The two recoil-energy bins are
\begin{align}
160<E_R<200~\keV,&\qquad n_1=0,\\
225<E_R<271~\keV,&\qquad n_2=1.
\end{align}
The expected count is
\begin{equation}
N_i=\mathcal E\int_{E_i^{\min}}^{E_i^{\max}}
\dd E_R\,\epsilon_{\rm LZ}(E_R)\frac{\dd R}{\dd E_R},
\label{eq:Ni}
\end{equation}
with $\mathcal E=2.84$ tonne-year. For the detector folding we use a smooth interpolation of the total nuclear-recoil signal-efficiency curve shown in Supplemental Fig.~S2 of Ref.~\cite{LZ2026}. The interpolation is constrained by the published behavior of that curve, including the approximately $96\%$ average signal efficiency between $14$ and $250~\keV$ and the $50\%$ efficiency point at $269.9~\keV$. This procedure is sufficient for the present broad-bin diagnostic, but it is not a replacement for the collaboration-level detector response in $(S1c,S2c)$ space. The numerical likelihood results below are therefore explicitly phenomenological and should not be interpreted as the official LZ profile likelihood.

The likelihood is
\begin{equation}
\cL=\prod_{i=1}^{2}\frac{N_i^{n_i}e^{-N_i}}{n_i!},
\label{eq:likelihood}
\end{equation}
and we define $\chi^2=-2\ln\cL$ and $\Delta\chi^2=\chi^2-\chi^2_{\min}$. For fixed masses and geometry, both $N_i$ scale as $\epsilon^2$. Thus the kinetic mixing can be profiled analytically. For $(n_1,n_2)=(0,1)$ the maximum with respect to the common normalization satisfies
\begin{equation}
N_1+N_2=1.
\label{eq:profileeps}
\end{equation}
This relation is useful because it separates the spectral information, encoded in $N_2/(N_1+N_2)$, from the overall portal normalization. Because the data consist of only one event in two broad bins and the present likelihood omits the full background and nuisance-parameter model, we use $\Delta\chi^2$ only as a relative likelihood diagnostic and do not assign asymptotic confidence-level coverage to the displayed contours. In particular, the reference levels $\Delta\chi^2=2.30$ and $6.18$ are shown only as convenient contour values and are not labeled as $68\%$ or $95\%$ confidence regions.

\section{Numerical results}
\label{sec:numerical_results}

\begin{figure}[t]
\centering
\includegraphics[width=0.95\columnwidth]{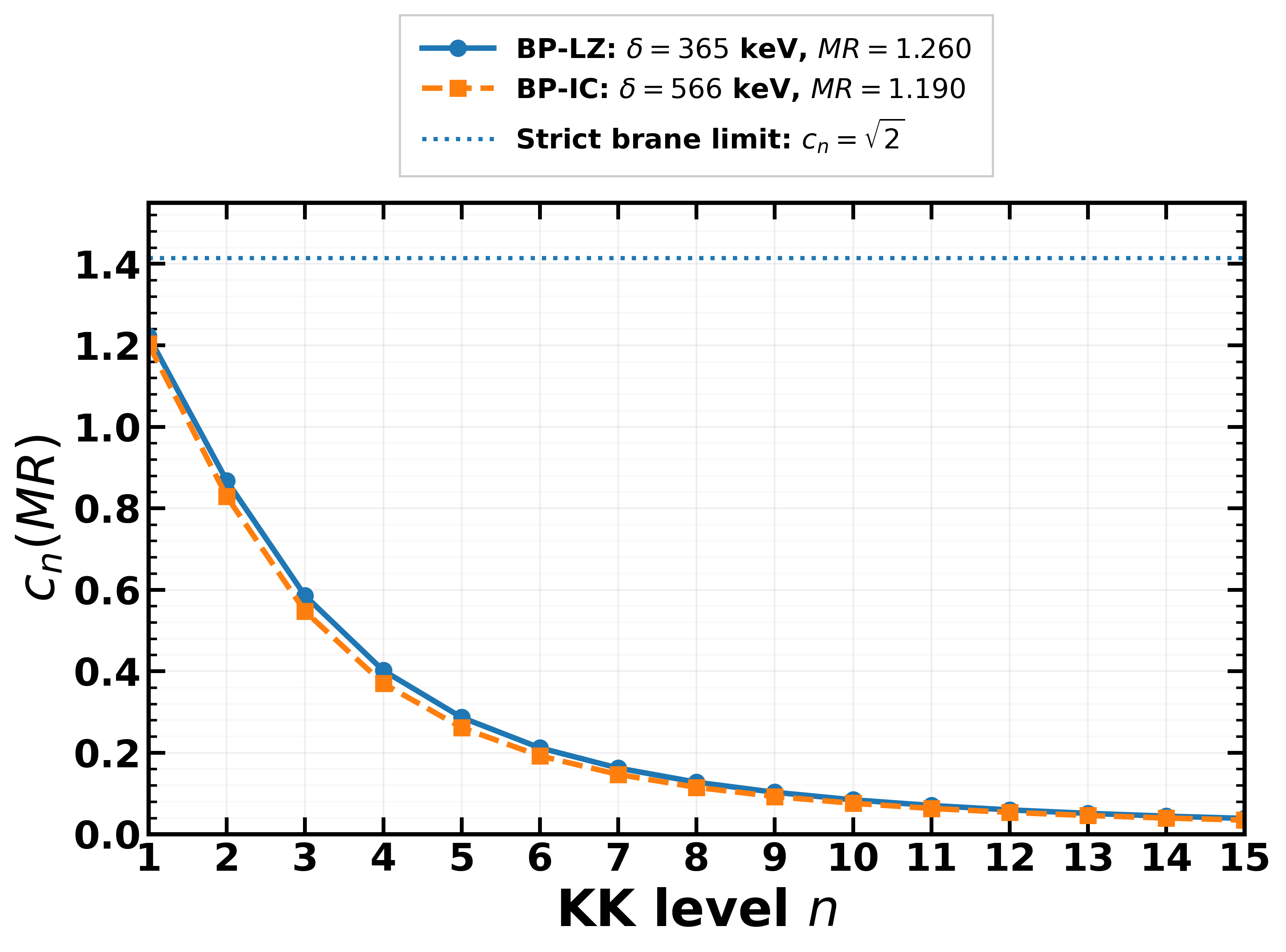}
\caption{Localization overlap coefficients $c_n(MR)$ for BP-LZ and BP-IC. The horizontal dotted line is the strict-brane value $\sqrt2$. The two curves are close because the benchmark localization parameters, $MR=1.260$ and $MR=1.190$, differ only moderately, but the higher KK levels are progressively filtered in both cases.}
\label{fig:overlaps}
\end{figure}

\begin{figure*}[t]
\centering
\includegraphics[width=0.94\textwidth]{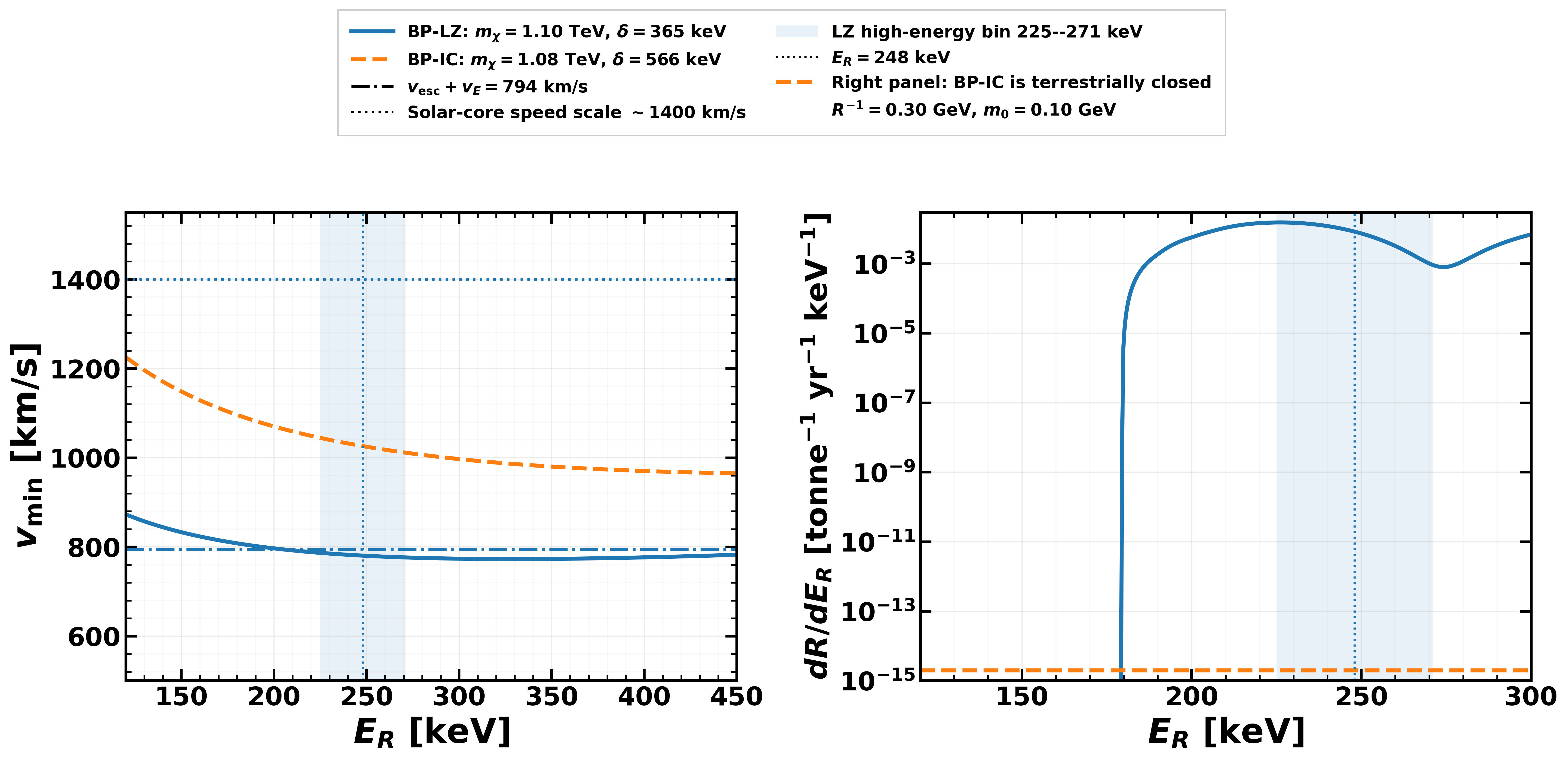}
\caption{Updated kinematic comparison for the two benchmarks. Left: $v_{\min}(E_R)$ in $^{131}\mathrm{Xe}$, showing BP-LZ close to the terrestrial speed ceiling $v_{\rm esc}+v_E$ while BP-IC lies above it; the $\sim1400~\mathrm{km/s}$ line is a characteristic solar-core speed scale. Right: the terrestrial xenon recoil spectrum. BP-LZ produces the high-recoil signal, whereas BP-IC is kinematically closed under the adopted halo model and is shown only as a dashed reference floor.}
\label{fig:kinematics-spectra}
\end{figure*}

\begin{figure*}[t]
\centering
\includegraphics[width=0.94\textwidth]{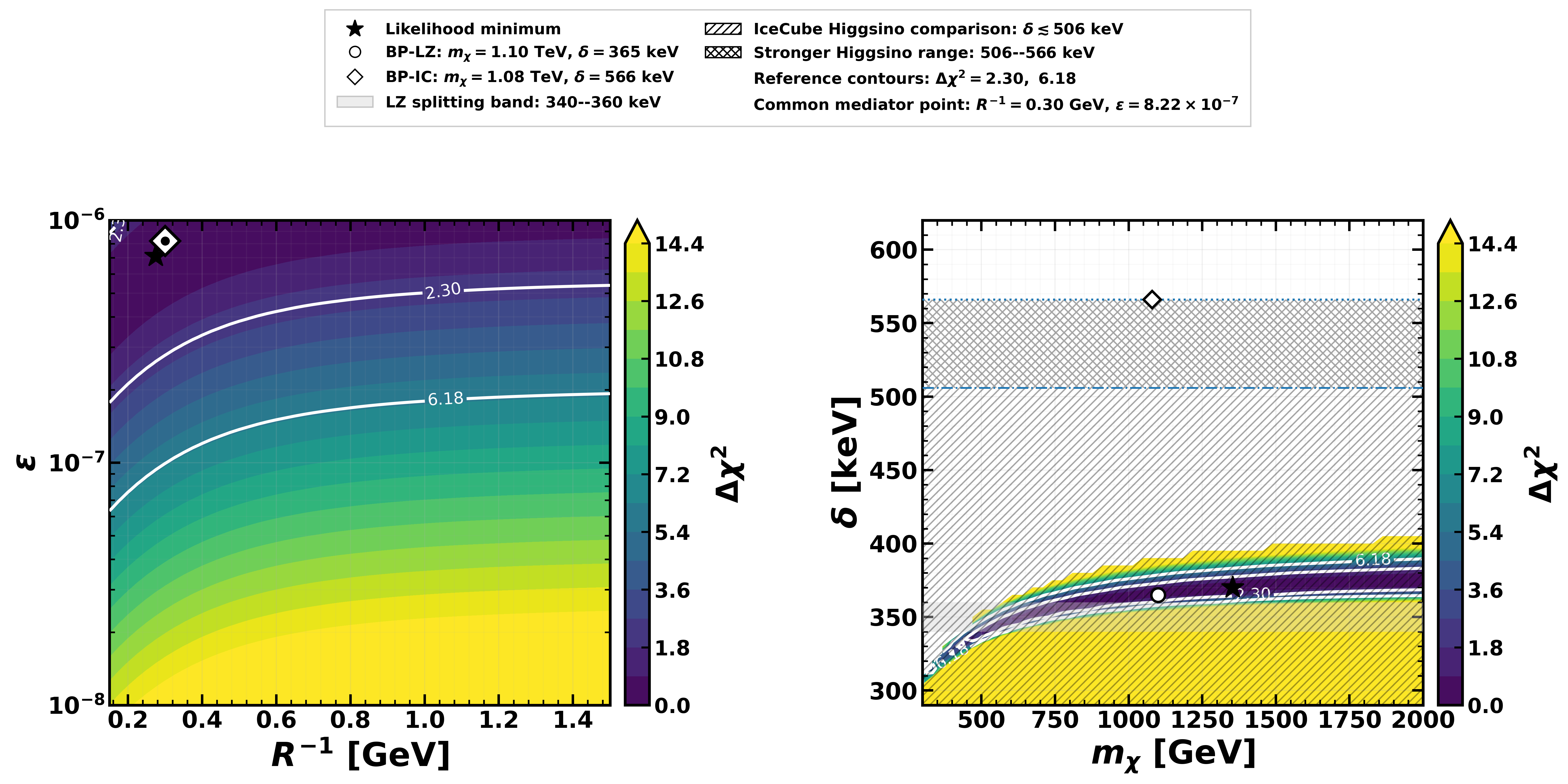}
\caption{Updated two-panel likelihood summary. Left: simplified BP-LZ likelihood in the $(R^{-1},\epsilon)$ plane. The star marks the grid minimum; the nested square/circle at the common mediator point indicates that BP-IC and BP-LZ use the same $(R^{-1},\epsilon)=(0.30~\GeV,8.22\times10^{-7})$. This does not represent a separate terrestrial LZ likelihood for BP-IC. Right: fixed-coupling LZ diagnostic in the $(m_\chi,\delta)$ plane, with BP-LZ and BP-IC overlaid. The gray strip marks the LZ splitting band, while the hatched regions show the Higgsino solar-capture thresholds $\delta\lesssim506~\keV$ and $506$--$566~\keV$ inferred from IceCube for comparison only. These IceCube bands are not direct exclusions of the present singlet KK dark-photon model. The displayed $\Delta\chi^2=2.30$ and $6.18$ curves are relative-likelihood reference contours, not confidence regions with assigned asymptotic coverage.}
\label{fig:likelihood-summary}
\end{figure*}

{\color{black}
We keep the graphical presentation compact but retain four complementary figures. Figure~\ref{fig:enhancement} displays both the compactification-scale dependence of the KK propagator and the direct BP-LZ--BP-IC comparison. In figure~\ref{fig:kinematics-spectra} we compare the terrestrial inelastic kinematics and xenon spectra. In figure~\ref{fig:overlaps} the localization filtering is made explicit through the first fifteen overlap coefficients. Finally, in figure~\ref{fig:likelihood-summary} we combine the $(R^{-1},\epsilon)$ LZ likelihood with the $(m_\chi,\delta)$ fixed-coupling diagnostic and the Higgsino-IceCube comparison bands. The Helm form factor, individual KK masses, and localization relation are not plotted separately because they are already summarized analytically and in the tables.
}

{\color{black}
We use two representative mass--splitting points,
\begin{align}
{\rm BP\mbox{-}LZ}:&\quad
m_\chi=1.10~\TeV,\qquad \delta=365~\keV,
\nonumber\\
{\rm BP\mbox{-}IC}:&\quad
m_\chi=1.08~\TeV,\qquad \delta=566~\keV,
\label{eq:twobenchmarks}
\end{align}
while the common mediator parameters are
\begin{equation}
{\color{black}
\begin{aligned}
R^{-1}&=0.30~\GeV, & m_0&=0.10~\GeV,\\
\alpha_D&=0.10, & \delta_{\rm UV}&=1~\GeV.
\end{aligned}
}
\label{eq:benchmark}
\end{equation}
The corresponding localization parameters are
$MR=1.260$ for BP-LZ and $MR=1.190$ for BP-IC. The second point is not an IceCube best fit to the present model; it is a comparison point chosen at the stronger Higgsino solar-capture threshold of Ref.~\cite{SolarCapture2026}. Profiling the two-bin likelihood over the kinetic-mixing normalization gives
\begin{equation}
\epsilon=8.22\times10^{-7}.
\end{equation}
The expected counts are
\begin{equation}
(N_{160-200},N_{225-271})=(0.124,0.876),
\label{eq:benchcounts}
\end{equation}
corresponding to $\chi^2=2.265$ for the two-bin statistic.
}

For orientation, Table~\ref{tab:benchmarkderived} collects several derived
quantities that connect the geometry directly to the recoil kinematics.
{\color{black}
\begin{table}[t]
\caption{Derived quantities for the two representative mass--splitting points.
The minimum speed is evaluated for a representative $^{131}\mathrm{Xe}$ nucleus
at $E_R=248~\keV$. BP-IC is an IceCube-motivated comparison point rather than a
direct IceCube exclusion of the present KK model.}
\label{tab:benchmarkderived}
\begin{ruledtabular}
\begin{tabular}{lcc}
Quantity & BP-LZ & BP-IC\\
\hline
$m_\chi$ [TeV] & $1.10$ & $1.08$\\
$\delta$ [keV] & $365$ & $566$\\
$MR$ & $1.260$ & $1.190$\\
$c_1$ & $1.223$ & $1.203$\\
$c_2$ & $0.868$ & $0.829$\\
$c_3$ & $0.585$ & $0.547$\\
$v_{\min}(248~\keV)$ [km/s] & $781$ & $1026$\\
terrestrial LZ reach & marginal & closed
\end{tabular}
\end{ruledtabular}
\end{table}
}

{\color{black}
In figure~\ref{fig:overlaps} we compare the localization overlap factors for the two benchmark splittings. 
The difference is modest but systematic: BP-LZ gives $(c_1,c_2,c_3)\simeq(1.223,0.868,0.585)$, while BP-IC gives $(1.203,0.829,0.547)$. Both sequences decrease rapidly for higher KK number, illustrating directly that finite localization filters the tower well before the strict-brane value $c_n=\sqrt2$ can be maintained over many modes.
}

At exactly the same $(\alpha_D,\epsilon,m_0)$ but with only a single dark photon, we obtain
\begin{equation}
(N_{160-200},N_{225-271})_{\rm single}=(0.0323,0.205),
\end{equation}
so the KK tower enhances the expected population of the high-energy bin by
\begin{equation}
\frac{N_{225-271}^{\rm KK}}{N_{225-271}^{\rm single}}=4.27.
\label{eq:enhanceN}
\end{equation}
{\color{black}This provides a quantitative indication that the extra dimension can be experimentally relevant even when the zero-mode kinetic mixing is very small.} It is important, however, to distinguish a fixed-coupling
rate enhancement from statistical evidence for the tower. After independently
profiling the common normalization, the relevant two-bin shape fractions are
\begin{equation}
p_2^{\rm KK}=0.876,\qquad
p_2^{\rm single}\simeq0.864.
\label{eq:shapeFractions}
\end{equation}
For the present $(0,1)$ data this corresponds to
$\Delta\chi^2_{\rm single-KK}\simeq0.03$ in the simplified two-bin statistic.
{\color{black}In this two-bin treatment, the discriminating information is therefore carried primarily by the finer recoil-energy dependence of the KK spectrum, which becomes increasingly testable as additional high-recoil events are accumulated.}

{\color{black}
In figure~\ref{fig:kinematics-spectra} we make the role of the two splittings explicit. BP-LZ remains barely accessible to the terrestrial high-velocity tail and produces the high-recoil spectrum of interest, whereas BP-IC is kinematically closed in xenon under the adopted Standard Halo Model. The latter point is therefore useful as a solar-capture comparison rather than as a second terrestrial LZ fit.
}

To isolate the shape information we define
\begin{equation}
\cR(E_R)=\frac{(\dd R/\dd E_R)_{\rm KK}}
{(\dd R/\dd E_R)_{\rm single}}.
\label{eq:ratio}
\end{equation}
The KK contribution is not equivalent to a constant rescaling of the
single-mediator rate. For $R^{-1}=0.30~\GeV$, the enhancement in the
high-recoil region is approximately a factor of four, but its magnitude
changes with recoil energy because the relative contributions of the KK
propagators depend on the momentum transfer $q=\sqrt{2m_AE_R}$. This
residual spectral dependence provides the more robust discriminator between a KK tower and a single mediator: although a change in $\epsilon$ can compensate the overall event rate, it cannot in general reproduce the energy-dependent distortion generated by the tower. With multiple high-energy recoil events, this spectral information could therefore be used to test the KK interpretation.

Before performing the scan, it is useful to identify the compactification regime to which the high-energy LZ event is intrinsically sensitive. Since a recoil near $E_R\simeq248~\keV$ in xenon corresponds to a momentum transfer $q\simeq0.24$--$0.25~\GeV$, the KK tower has its largest spectroscopic impact when the level spacing is comparable to this momentum,
\begin{equation}
R^{-1}\sim q_{\rm LZ}\sim {\cal O}(0.1\text{--}1)~\GeV .
\label{eq:scale_matching}
\end{equation}
This observation explains why the sub-GeV compactification region is the natural target of the present analysis. If $R^{-1}\gg q_{\rm LZ}$, the excited modes satisfy $m_n^2\simeq n^2/R^2\gg q^2$ and their propagators decouple, so the theory approaches the single-dark-photon limit. In the opposite regime, $R^{-1}\ll q_{\rm LZ}$, many closely spaced KK states contribute coherently and the rate can be strongly enhanced; however, the interpretation then becomes increasingly sensitive to the five-dimensional cutoff, boundary-localized operators, and tower-level experimental constraints. The intermediate regime $R^{-1}\sim q_{\rm LZ}$ is therefore both phenomenologically efficient and theoretically transparent.

We next fix the following values
\[(m_\chi,\delta,m_0,\alpha_D) =(1.1~\TeV,365~\keV,0.10~\GeV,0.10),\]
and scan the compactification scale $R^{-1}$ together with the kinetic
mixing $\epsilon$. Decreasing $R^{-1}$ reduces the spacing between
successive KK modes, allowing more of the tower to contribute at the
momentum transfers relevant for the high-recoil LZ event. The resulting
enhancement of the scattering amplitude implies that a smaller value of
$\epsilon$ is required to reproduce a fixed event yield. Consequently,
$R^{-1}$ and $\epsilon$ exhibit a pronounced correlation: a denser and
lighter KK spectrum can be compensated by weaker kinetic mixing.

Throughout this scan, we keep $\delta$ and $\delta_{\rm UV}$ fixed and therefore retain the value  $MR=1.260$ fixed. Consequently, varying $R^{-1}$ implicitly corresponds to varying the physical bulk localization mass according to $M=(MR)R^{-1}$. This prescription isolates the mediator-tower dependence on the compactification scale while preserving the chosen inelastic splitting. Likewise, $\epsilon$ is scanned as an effective four-dimensional parameter; through Eq.~\eqref{eq:mixingmodes}, different points in the plane correspond to the appropriate microscopic brane-mixing coefficient $\epsilon_5$ for the chosen $R$ and $\Lambda_5$.

{\color{black}
The corresponding two-bin likelihood is shown in the left panel of Fig.~\ref{fig:likelihood-summary}. The grid minimum occurs at
\begin{equation}
R^{-1}=0.296~\GeV,\qquad
\epsilon=8.23\times10^{-7},
\end{equation}
with
\begin{equation}
(N_1,N_2)=(0.127,0.893),
\qquad \chi^2_{\min}=2.266.
\end{equation}
The close agreement with the reference benchmark reflects the deliberately chosen $R^{-1}=0.30~\GeV$. More importantly, the likelihood exhibits an extended degeneracy direction: a denser KK tower can be compensated by a smaller zero-mode mixing.

The BP-IC marker is also shown at the same mediator coordinates as BP-LZ because both benchmarks adopt the common values $R^{-1}=0.30~\GeV$ and $\epsilon=8.22\times10^{-7}$. The nested marker is only a bookkeeping device for the shared mediator benchmark; BP-IC itself has no conventional terrestrial LZ likelihood surface because the $566~\keV$ transition is kinematically closed in xenon under the adopted halo model.
}

This degeneracy is physically instructive. Two broad recoil bins constrain mainly an overall rate and one coarse spectral ratio. A true determination of $R^{-1}$ would require more spectral information, a second nuclear target, or independent constraints on $\epsilon$. The residual energy dependence of the KK-to-single-mediator ratio shows that such a determination is possible in principle.

The likelihood minimum at $R^{-1}\simeq0.30~\GeV$ should therefore not be interpreted as a measurement of the compactification scale from a single event. Rather, it identifies the scale-matching region in which the first KK excitations lie close to the characteristic momentum transfer of the LZ recoil. The physically robust result is the existence of a correlated band in the $(R^{-1},\epsilon)$ plane together with a residual energy-dependent distortion that cannot be removed by a single rescaling of $\epsilon$.

{\color{black}
Finally, we examine a complementary fixed-coupling slice in the $(m_\chi,\delta)$ plane. We keep $R^{-1}=0.30~\GeV$, $m_0=0.10~\GeV$, $\alpha_D=0.10$, and $\epsilon=8.22\times10^{-7}$ fixed, while the localization parameter is recalculated at each splitting through Eq.~\eqref{eq:MRinverse}. This scan is shown in the right panel of Fig.~\ref{fig:likelihood-summary}. Because the normalization is not reprofiled at every point, this panel should be read as a benchmark diagnostic of the combined kinematic and localization dependence rather than as a model-independent confidence region.
}

On the finite grid the minimum of this fixed-coupling slice lies near
\begin{equation}
m_\chi\simeq1.42~\TeV,\qquad \delta\simeq370~\keV,
\end{equation}
where the expected counts are approximately $(0.095,0.813)$ and $\chi^2\simeq2.23$. The reference $1.1~\TeV$, $365~\keV$ point is therefore already close to this minimum. Given the simplified likelihood, the fixed normalization in this scan, and the single observed high-energy event, we do not attach physical significance to the exact grid minimum. The robust statement is that splittings near the kinematic boundary strongly reduce the lower-energy population while the KK tower can maintain an observable high-energy rate.

{\color{black}
The two benchmark markers in Fig.~\ref{fig:likelihood-summary} sharpen this
interpretation. BP-LZ sits in the terrestrial edge region, while BP-IC is far
above the xenon kinematic reach. This difference is not a statement that
IceCube excludes BP-IC in our model; rather, it illustrates why the Sun can test
splittings that a terrestrial xenon experiment cannot. A model-specific
IceCube prediction would require the KK-mediated solar capture rate together
with the annihilation and decay pattern of the dark-sector states.
}

\section{Phenomenological implications and limitations}

The use of bulk dark photons and brane kinetic mixing is well established \cite{Rizzo2018I,Rizzo2018II,RizzoWojcik2021}. The present construction differs in emphasis from those studies in two ways. First, the observable of interest is a high-energy endothermic nuclear recoil rather than thermal freeze-out or low-energy accelerator production. Second, the fermion is finitely localized, so the same parameter $MR$ that produces the pseudo-Dirac splitting also suppresses higher KK couplings according to Eq.~\eqref{eq:cn}. This yields a geometry-dependent spectral filter that is absent if dark matter is treated as an exactly brane-localized point particle.

The recently proposed naturally resonant extra-dimensional dark-matter scenario of Lee and Tsai also employs fermionic dark matter and a dark photon on an orbifold and demonstrates distinctive direct-detection and accelerator signatures \cite{LeeTsai2026}. Our setup is complementary: we do not rely on an annihilation resonance. Instead, the relevant scales are selected by the momentum transfer in a high-energy direct-detection event, and the distinctive structure is the localization-filtered $t$-channel KK sum.

{\color{black}
In the present singlet-dark-matter realization, the bulk dark photon provides an additional observable handle: the compactification radius controls the spacing of mediator poles, while the localization profile controls their residues. Consequently, the same geometry that generates the inelastic threshold can leave a momentum-dependent imprint on the recoil spectrum. This mediator-spectroscopy aspect is the principal new element of the present analysis.
}

{\color{black}
For orientation, the benchmark value $R^{-1}=0.30~\GeV$ corresponds to $R\simeq0.66~{\rm fm}$, or an interval length $\pi R\simeq2.07~{\rm fm}$. This comparatively low compactification scale is selected by the momentum-transfer regime of the LZ high-recoil event, $q_{\rm LZ}\sim0.25~\GeV$, rather than by the conventional TeV-scale extra-dimensional hierarchy problem.
}

The benchmark zero-mode mixing, $\epsilon\sim8\times10^{-7}$ at $m_0=100~\mathrm{MeV}$, lies in a region where existing dark-photon constraints are potentially relevant \cite{DarkPhotonReview}. A direct one-to-one application of a single-mediator exclusion curve is not generally justified here, however, because the physical signal involves a KK tower. Excited states have different masses and brane wave functions, and their branching fractions depend on the dark spectrum. Precision observables, beam-dump experiments, meson decays, and missing-energy searches can therefore provide complementary information on $R^{-1}$ and $\epsilon$.

{\color{black}
Solar capture is another important complementary probe. In the Higgsino case,
tree-level EW scattering on heavy solar elements leads to efficient
capture and, after annihilation to $W^+W^-$ and $ZZ$, a high-energy neutrino
signal. Using ten years of IceCube solar-neutrino data, Ref.~\cite{SolarCapture2026}
finds a splitting threshold of approximately $506~\keV$ from tree-level capture
alone and up to $566~\keV$ when the captured population thermalizes efficiently.
The same numerical bound cannot be imported into our model: the dark matter is
a Standard-Model singlet, the capture amplitude is controlled by the
localization-filtered dark-photon KK propagator, and annihilation may proceed
predominantly into light dark-sector vectors rather than directly into
EW gauge bosons. A dedicated solar analysis of the present model is
therefore left for future work.
}

The benchmark scattering amplitude is dominated by the first few KK states:
$m_1\simeq0.316~\GeV$, $m_2\simeq0.608~\GeV$, and
$m_3\simeq0.906~\GeV$, while their dark-matter overlap factors decrease as
$c_1\simeq1.22$, $c_2\simeq0.87$, and $c_3\simeq0.59$. Consequently, the
low-energy prediction does not require a physical tower extending to
$n=100$; that value is only a numerical convergence parameter. The EFT result
is stable provided the physical cutoff lies sufficiently above the first few
modes that dominate the momentum range of interest.

Boundary-localized gauge kinetic terms can also modify the KK spectrum and couplings. Such terms play an important role in consistent five-dimensional kinetic-mixing constructions \cite{Rizzo2018I,Rizzo2018II}. We have intentionally omitted them in order to isolate the leading localization effect. Their inclusion is a natural next step once a larger direct-detection data set justifies a more detailed spectral analysis.

Endothermic scattering at $\delta\sim365~\keV$ probes the extreme high-speed tail of the local dark-matter distribution. Consequently, the absolute rate is more sensitive to halo assumptions than standard elastic WIMP scattering, a point emphasized in recent analyses of the same LZ event \cite{McCabe2026,BaerBarger2026,Yuan2026}. We therefore do not use the absolute normalization as a precision determination of the geometry. A future detector-level analysis should vary the escape speed, Earth speed, local density, and possible non-Maxwellian components, and should test the associated annual modulation. The more robust target of the present construction is the energy dependence of the KK enhancement at fixed geometry.

At recoil energies near $250~\keV$, the momentum transfer is also large enough that nuclear-response and detector-response details matter. The Helm form factor and the one-dimensional efficiency folding used here are adequate for a transparent comparison of mediator structures, but a collaboration-level recast would require the nuclear responses and detector simulation used by LZ, together with the full signal and background probability densities in $(S1c,S2c)$ space. The present two-bin likelihood follows the simplified strategy of Ref.~\cite{Su2026}; it is deliberately not presented as the official LZ profile likelihood.

Heavy nuclei are favored by endothermic kinematics. Tungsten therefore provides a particularly useful cross-check, as emphasized in Ref.~\cite{Su2026}. A future CRESST high-energy analysis could test whether a xenon feature is associated with the same inelastic transition. In the present model the comparison becomes even more informative because the momentum transfer differs between xenon and tungsten, sampling the KK propagator at different $q$. A combined xenon--tungsten analysis could therefore help break the $R^{-1}$--$\epsilon$ degeneracy visible in the left panel of Fig.~\ref{fig:likelihood-summary}.

A single event can always be mimicked by changing an interaction normalization. The distinctive prediction of the extra-dimensional model is therefore not the existence of a point in parameter space with $N\sim1$, but the energy dependence of the tower enhancement. If future LZ exposure reveals several events across the extended recoil window, one can compare a single-mediator hypothesis against the correlated shape
\begin{equation}
\cR(E_R)=\left|\frac{\cP_{\rm loc}(q(E_R))}{\cP_0(q(E_R))}\right|^2.
\end{equation}
Because $c_n$ is fixed by the same $MR$ that controls $\delta$, the shape is not arbitrary. This correlation is the main route by which direct detection could become a probe of dark-sector geometry rather than merely a measurement of an effective cross section.

\section{Conclusions}

We have developed a five-dimensional realization of inelastic dark matter in which a pseudo-Dirac fermion scatters through a bulk dark-photon tower. The central feature is geometric economy: exponential localization toward one boundary suppresses a distant-brane Majorana operator, generating
\begin{equation}
\delta=\delta_{\rm UV}e^{-2\pi MR},
\end{equation}
while the same finite-width profile determines the coupling to each gauge KK mode through Eq.~\eqref{eq:cn}. The result is a localization-filtered mediator propagator rather than a universal KK sum.

For an exactly brane-localized interaction the dark-photon tower sums to the compact analytic form $\cP_{\rm brane}=\pi R\coth(\pi RQ)/Q$. Finite fermion localization suppresses higher modes, but substantial enhancement remains when the momentum transfer is comparable to the compactification scale. This makes the high-energy LZ recoil window particularly interesting: $E_R\sim250~\keV$ in xenon corresponds to $q\sim0.25~\GeV$, directly probing $R^{-1}$ in the sub-GeV to GeV regime.

{\color{black}
At the LZ benchmark
$m_\chi=1.10~\TeV$, $\delta=365~\keV$, $R^{-1}=0.30~\GeV$,
$m_0=0.10~\GeV$, and $\alpha_D=0.1$, the splitting corresponds to
$MR=1.260$. Profiling the two-bin Poisson likelihood gives
$\epsilon=8.22\times10^{-7}$ and expected counts $(0.124,0.876)$.
At the same microscopic coupling, the high-energy rate is larger by a factor
$4.27$ than for a single dark photon. The comparison point
$m_\chi=1.08~\TeV$, $\delta=566~\keV$ instead has $MR=1.190$ and requires
$v_{\min}(248~\keV)\simeq1026~\mathrm{km/s}$, placing it beyond the terrestrial
LZ kinematic reach but within the qualitative velocity regime relevant for
solar capture. This two-point comparison makes clear that the LZ and IceCube
observables probe complementary velocity environments rather than identical
parameter exclusions. {\color{black}The key discriminator is the energy-dependent spectral distortion generated by the localization-filtered KK tower, which can be tested directly as the high-recoil spectrum becomes better populated.}
}

\appendix
\section{Localization-factor derivation}
\label{app:cn}

For the normalized profile $|f_\chi(y)|^2\propto e^{-2My}$ and the gauge mode $f_n^A(y)\propto\cos(ny/R)$, the ratio of the excited-mode overlap to the zero-mode overlap is
{\color{black}
Defining the normalized overlap ratio
\begin{align}
\mathcal I_n&\equiv
\frac{\int_0^{\pi R}\dd y\,e^{-2My}\cos(ny/R)}
{\int_0^{\pi R}\dd y\,e^{-2My}},\nonumber\\
\mathcal I_n&=
\frac{(2MR)^2[1-(-1)^ne^{-2\pi MR}]}
{[(2MR)^2+n^2][1-e^{-2\pi MR}]}.
\end{align}
}
Multiplication by the $\sqrt2$ normalization of the excited gauge mode gives Eq.~\eqref{eq:cn}.

For completeness, the strict-brane expression in
Eq.~\eqref{eq:Pbrane} follows from
\begin{equation}
\sum_{n=-\infty}^{\infty}\frac{1}{n^2+a^2}
=\frac{\pi}{a}\coth(\pi a),
\label{eq:cothidentity}
\end{equation}
with $a=RQ$. Splitting the $n=0$ term from the symmetric positive and negative
modes immediately gives Eq.~\eqref{eq:Pbranesum}. This identity is also a
useful numerical cross-check because the direct unfiltered KK sum must approach
the analytic result independently of the localization calculation.

The localized series is even better behaved in the ultraviolet. Combining
Eqs.~\eqref{eq:cnasymptotic} and \eqref{eq:Ploc} shows that its large-$n$
remainder scales as $\sum_{n>N}n^{-4}\sim(3N^3)^{-1}$. Consequently, the
choice $n_{\max}=100$ is far more than is required for numerical convergence
of the benchmark observables. The physically meaningful restriction is instead
that the KK modes retained in a five-dimensional effective description lie
below the ultraviolet cutoff $\Lambda_5$; the phenomenology discussed in the
main text depends primarily on the first few modes listed in
Table~\ref{tab:kkmodes}.

\end{document}